\documentclass[%
 reprint,
showpacs,
showkeys,
preprintnumbers,
 amsmath,amssymb,
 aps,
prc,
]{revtex4-2}

\usepackage{graphicx}
\usepackage{amsmath}
\usepackage{dcolumn}
\usepackage{bm}
\usepackage{xcolor}
\usepackage[normalem]{ulem}

\graphicspath{{./}{./figs/}}

\begin{document}

\preprint{APS/123-QED}

\title{Indirect Measurement of the $\rm S^*(E)$ Factor for $\rm {}^{12}C({}^{12}C,\mathit{p}){}^{23}Na$ at Gamow Energies via the Trojan Horse Method with Near-$0^{\circ}$ Spectator Detection}
\thanks{Supported by Natural Science Foundation of Beijing Municipality (1222022) and National Natural Science Foundation of China (12075031, 12275360)}%


\author{Chengbo Li$^1$}
 \email{lichengbo@bjast.ac.cn; lichengbo2008@163.com}
\author{Huiming Jia$^2$}
 \email{jiahm@cnncmail.cn}
\author{Qungang Wen$^3$}
 \email{qungang@ahu.edu.cn}
\author{Chengjian Lin$^2$}
\author{Lei Yang$^2$}
\author{Feng Yang$^2$}
\author{Nanru Ma$^2$}
\author{Tianpeng Luo$^2$}
\author{Xuepeng Sun$^1$}
\author{Shangkun Shao$^1$}
\author{Xuejian Wang$^3$}

\affiliation{ 1. Institute of Radiation Technology, Beijing Academy of Science and Technology, Beijing 100875, China.}
 
\affiliation{2. Department of Nuclear Physics, China Institute of Atomic Energy, Beijing 102413, China}

\affiliation{3. School of Physics and Materials Science, Anhui University,  Hefei 230601, China}


\date{\today}

\begin{abstract}
The astrophysical $S^*(E)$ factor for the $^{12}$C+$^{12}$C reaction within the Gamow window ($ E_G=1.5\pm 0.3$ MeV) plays a pivotal role in modeling stellar carbon burning and explosive nucleosynthesis scenarios.
However, direct measurements — or even simple extrapolations — at these energies are severely hindered by Coulomb suppression and the possible presence of narrow resonances.
To address this challenge, we performed an indirect measurement of the  $\rm {}^{12}C(^{16}O,\alpha \mathit{p}){}^{23}Na$ reaction at the HI-13 Tandem Accelerator, employing $\rm {}^{16}O=({}^{12}C \oplus \alpha)$ as the Trojan Horse nucleus.
A key innovation of this Trojan Horse Method (THM) study is the implementation of a copper beam-stopper foil, which enabled the detection of spectator particles near $0^{\circ}$  — the angular region where their yield is maximized under quasi-free kinematics.
The $S^*(E)$  factor for the $\rm {}^{12}C({}^{12}C,\mathit{p}){}^{23}Na$ reaction in the astrophysically relevant energy range was extracted using the THM formalism based on the distorted-wave Born approximation (DWBA).
Our results confirm the presence of resonant structures within the Gamow window around 1.5 MeV in both the $p_0$ and $p_1$ proton channels. 
No evidence of a hindrance effect is observed in the measured energy range.
Without considering the resonance details, the overall trend of our results is qualitatively in reasonable agreement with the THM-Tumino2018 and TTIK2025 data, but differs significantly from the trend of the Modified-THM-Muk2019 data.

\end{abstract}


%
\maketitle


\section{\label{sec:intro} Introduction}

The $\rm ^{12}C+{}^{12}C$ reaction at astrophysical energies is important for understanding the carbon burning process in massive stars \cite{RMP2014} and explosive astrophysical scenarios such as Type Ia supernovae and X-ray bursts \cite{RMP2002, Ast2001}. 

The temperature for typical stellar carbon burning is approximately 0.8 GK, corresponding to a Gamow window centered at $ E_G=1.5 \pm 0.3$ MeV. 
For the extended temperature range of 0.6–1.2 GK, the corresponding energy range is 1–3 MeV.
However, this energy lies far below the Coulomb barrier ($\approx$ 7.8 MeV).
At energies below approximately 2.6 MeV, the carbon fusion reaction proceeds predominantly through the $\alpha$ and $p$ emission channels \cite{Lyj2020}.

The $\rm ^{12}C+{}^{12}C$ reaction has been studied since the 1960s \cite{Bro1960, Alm1960}, revealing unexpected resonance structures.
These discoveries spurred extensive efforts to measure the reaction cross-section near astrophysical energies using diverse techniques, including charged-particle spectroscopy \cite{Pat1969, Maz1973, Bec1981, Zic2018}, $\gamma$-ray detection \cite{Hig1977, Ket1980, Das1982, Agu2006, Bar2006, Spi2007}, and coincidence measurements of charged-particle and $\gamma$-ray to suppress backgrounds \cite{Jcl2018, Fru2020, Tan2020}.

However, direct measurement of the $\rm ^{12}C+{}^{12}C$ reaction becomes increasingly challenging at lower energies due to the Coulomb barrier.
The lower limit of direct measurements is approximately 2.1 MeV, leaving the Gamow window largely unexplored.
Spillane \cite{Spi2007} found a strong resonance near $ E_{cm}$=2.14 MeV on the high-energy tail of the Gamow peak. This would increase the non-resonant reaction rate by a factor of five for stars near T=0.8 GK.

Concurrently, numerous theoretical models have been developed to understand the underlying reaction mechanism \cite{CF1988, Hin2007, Cop2009, AMD2021, DIM2024}.  
The $\rm ^{12}C+{}^{12}C$ system exhibits a quasimolecular structure, and the presence of corresponding resonances can enhance the cross section by several orders of magnitude. 
However, extrapolating the direct data to lower energies is unreliable due to the potential presence of resonances and significant discrepancies between the trends predicted by various theoretical models \cite{CF1988, Hin2007, Cop2009, AMD2021, DIM2024}, as illustrated in Figs.~\ref{figSp0} and~\ref{figSp1}.

Tumino et al. \cite{Tum2018} reported strong resonances between 0.8 and 2.7 MeV via the Trojan horse method (THM), especially near 1.5 MeV within the Gamow window.
However, their results have been questioned regarding the validity of the plane-wave impulse approximation (PWIA), the assumed spectator momentum distribution, and the reported resonance strength below 1 MeV  \cite{Muk2019,Muk2022,Txd2019,Znt2020,Bec2020}. 
All the arguments need to be tested by further experimental measurements, using either direct or indirect methods.

In the present work, we employ the Trojan Horse method based on DWBA instead of PWIA.

The binding energy of the Trojan Horse nucleus ${}^{14}\mathrm{N} = ({}^{12}\mathrm{C}\oplus d)$ is relatively high (10.27 MeV), and the spectator $d$ has a very small binding energy (2.225 MeV). This leads to a competing cluster structure, ${}^{14}\mathrm{N} = ({}^{13}\mathrm{C}\oplus p)$, which is more prone to breakup (7.55 MeV).

In this experiment, we chose $^{16}\mathrm{O}=( {}^{12}\mathrm{C} \oplus \alpha)-7.16$ MeV as the Trojan Horse nucleus \cite{a16O1984, a16O1988} because of its lower binding energy and the fact that the spectator $\alpha$-particle is tightly bound. Therefore, it is expected to more readily produce a quasi-free breakup.

The use of $\alpha$ as a spectator particle has been validated in previous THM experiments with the Trojan horse nucleus $\rm ^{6}Li= (\alpha \oplus \mathit{d})$ \cite{APJ1996,PRC2001,lcb2015,lcb2017} or $\rm ^{9}Be= (\alpha \oplus {}^{5}He)$ \cite{EPJA2020,EPJA2021}, with no evidence that it causes significant distortion.

A key innovation of our experiment is the use of a copper beam-stopper foil, enabling, for the first time in THM, the measurement of spectators around $0^\circ$, where most of the spectator particles are distributed under the quasi-free mechanism.

The $S^*(E)$ factor of $\rm {}^{12}C({}^{12}C,\mathit{p}){}^{23}Na$ in the Gamow energy region is extracted from indirect measurement of the three-body reaction $\rm {}^{12}C(^{16}O,\alpha \mathit{p}){}^{23}Na$.

\section{\label{sec:method} Trojan Horse Method}

\begin{figure}
\begin{center}
\includegraphics[width = 0.50\textwidth]{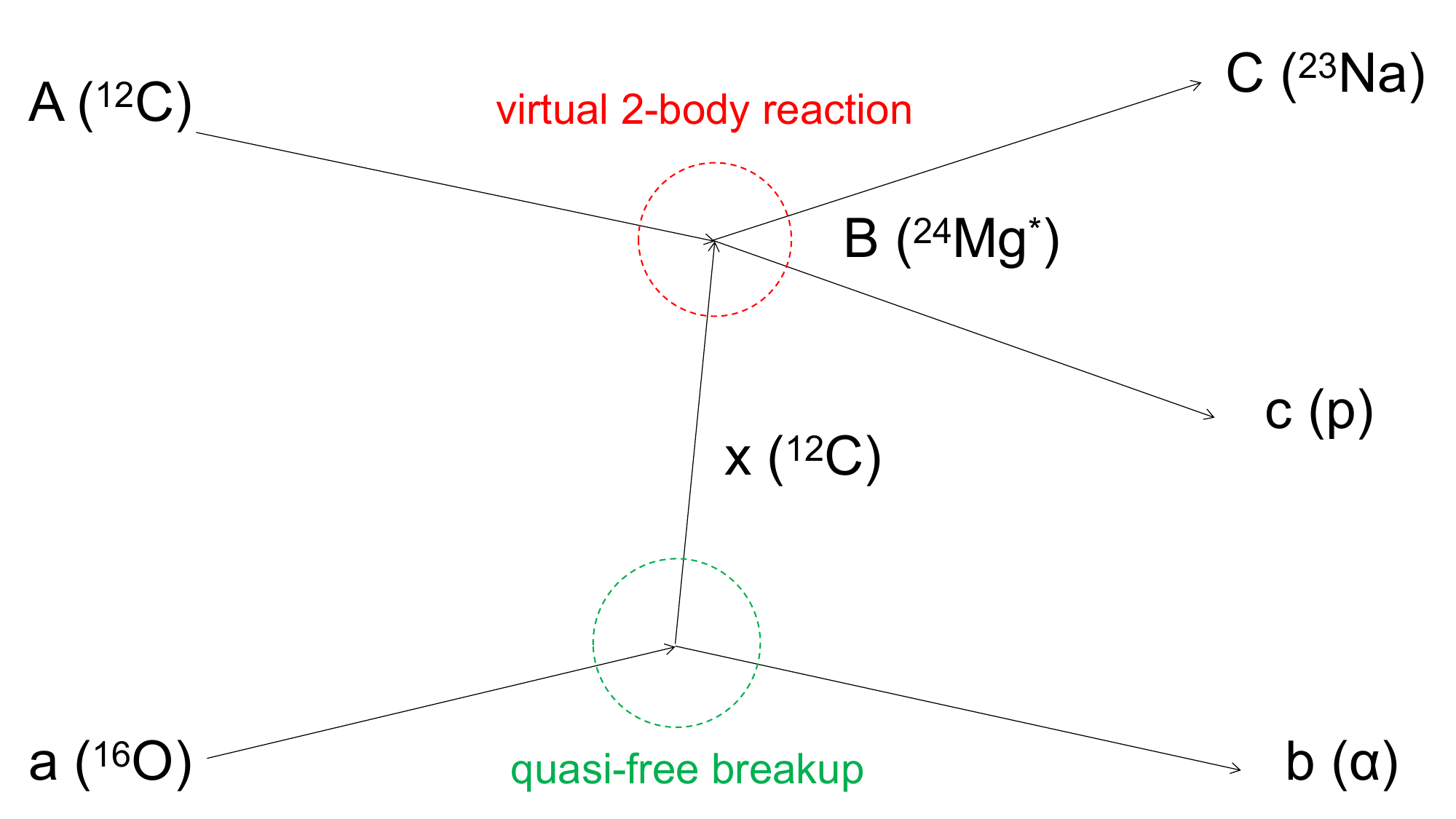}\\
\caption{(Color online) The schematic representation of Trojan horse method. General: A(a,bc)C; this work:  $\rm {}^{12}C(^{16}O,\alpha \mathit{p}){}^{23}Na$ }
\label{figTHM} 
\end{center}
\end{figure}

The Trojan horse method (THM) is a powerful indirect technique in experimental nuclear astrophysics, developed to determine the energy dependence of S factors of astrophysically relevant two-body reactions at very low energies, overcoming the influence of Coulomb barrier and electron screening effect \cite{Bau1986, Typ2003, Tri2014, Ber2018, epjrev2019}.
It has been successfully applied to numerous astrophysically important reactions \cite{epjrev2019, prl2007, prl2012, plb2015, apj2017, Tum2018, wen2008, wen2011, lcb2015, lcb2015a, wen2016, lcb2017,  wang2024}.

Initially, THM relied on the plane-wave impulse approximation (PWIA) \cite{Bau1986}. Subsequently, S. Typel and G. Baur enhanced the method using the distorted wave Born approximation (DWBA) framework to consider the influence of the Coulomb force on the wave function \cite{Typ2003}.
The basic assumptions of the DWBA based THM theory and the detailed theoretical derivation of the formalism employed can be found in Ref.~\cite{Typ2003}.

As shown in Fig. {\ref{figTHM}}, the THM is based on the quasi-free (QF) reaction mechanism, which allows one to derive indirectly the cross section at low energies of a 2-body reaction between charged particles $A+x \rightarrow C+c$ from the measurement of a suitable 3-body process $A+a \rightarrow C+c+b$ under the quasi-free kinematic conditions at high energy above the Coulomb barrier.
The Trojan horse nucleus $a=(x \oplus b)$ is considered to be dominantly composed of clusters $x$ and $b$.
Following the quasi-free breakup of $a$ induced by the interaction with $A$, the two-body reaction $A+x \to C+c$ proceeds virtually, while the spectator $b$ retains almost its original momentum distribution.

The Trojan-horse particles, in which the $s$-wave configuration dominates their ground states, exhibit a momentum distribution that peaks near zero. This characteristic defines the so-called quasi-free condition. Within this specific region of the three-body phase space, the cross-section for quasi-free reactions reaches its maximum. Based on this condition, by applying the law of conservation of energy, an energy relationship can be derived as follows \cite{Typ2003}:
\begin{equation}\label{eq:EqEAxqf}
    E_{Ax}^{qf}=E_{Aa}\left(1-\frac{\mu_{Aa}}{\mu_{Bb}}\frac{\mu_{bx}^{2}}{m_x^2}\right)-\varepsilon_{a}
\end{equation}
The entrance energy $E_{Aa}$ of the 3-body reaction is chosen above the Coulomb barrier to avoid the reduction in cross section. 
However, the energy $E_{Ax}$ of the 2-body reaction can be relatively small, reaching $E_G$ or sub-threshold energy region. This is mainly because the energy $E_{Aa}$ is partially used to supply the binding energy $\varepsilon_a$ of $a=(x\oplus b)$, and the Fermi motion of $x$ inside $a$ is used to span the energy region around the quasi-free energy point: $E_{Ax}=E_{Ax}^{qf} \pm E_{xb}$.
In the experiment, $E_{Ax}$ is inversely derived from the outgoing particle measurement: $E_{Ax}=E_{Cc}-Q_2$.

The basic theory is developed starting from a post-form distorted wave Born approximation of the T-matrix element. In the surface approximation the cross-section of the three-body reaction can be related to the S-matrix elements of the two-body reaction. In a modified plane wave approximation the relation between the two-body and three-body cross-sections becomes very transparent. 
The three-body reaction cross-section is decomposed into the product of three parts, the relationship is finally obtained as follows\cite{Typ2003}:
\begin{equation}\label{eq:sec3all}
    \frac{d^3\sigma}{dE_{Cc}d\Omega_{Cc}d\Omega_{Bb}} = K_F \cdot |W|^2  \cdot \frac{d\sigma^{TH}}{d\Omega}
\end{equation}

Where the first part $K_F$ is a kinematic factor containing the final-state phase-space factor.
$$K_F=\frac{\mu_{Aa}\mu_{Bb}\mu_{Cc}}{(2\pi)^5\hbar^6} \frac{k_{Bb}k_{Cc}}{k_{Aa}} \frac{16\pi^2}{k^2_{Ax}Q^2_{Aa}} \frac{v_{Cc}}{v_{Ax}}$$
with $\vec{Q}_{Aa}=\vec{k}_{Aa}- \frac{m_A}{m_x + m_A} \vec{k}_{Bb}$.

The second part $|W|^2$ is the momentum distribution of the spectator $b$ inside $a$.
$$W=-(\varepsilon_a + \frac{\hbar^2 Q^2_{Bb}}{2\mu_{bx}})\langle \exp(i\vec{Q}_{Bb}\cdot\vec{r}_{xb})\phi_x \phi_b |\phi_a \rangle$$
with $\vec{Q}_{Bb}=\vec{k}_{Bb}- \frac{m_b}{m_x + m_b} \vec{k}_{Aa}$.

The third part is the so-called two-body Trojan-Horse cross section, which is connected to the actual two-body reaction cross-section as follows:
$$ \frac{d\sigma^{TH}}{d\Omega}= T_l  \cdot \frac{d\sigma_l}{d\Omega}({Ax\rightarrow Cc)} $$
for $l$ wave (normally, only s-wave is used, $l$=0). Where the factor $T_l$ is defined as:
$$ T_l \rightarrow k^2 R^2 [F_l^2 + G_l^2] z_l^2 $$
where $F_l$ and $G_l$ are the regular and irregular Coulomb wave functions, and $z_l$ is the Riccati–Bessel function. 
The factor $T_l$ serves as a compensation for the Coulomb penetration effect (with the standard definition $P_l = k R / [F_l^2 + G_l^2]$). 

The energy relationship in Eq.(\ref{eq:EqEAxqf}) guides the selection of the incident beam energy ($E_{Ax}^{qf}$ is usually set to $E_G$), while the cross-section relationship in Eq.(\ref{eq:sec3all}) allows for the extraction of the relevant 2-body reaction cross section at low energies from the measured 3-body cross section after selecting quasi-free events.
Then, the  $S^*(E)$ factor can be determined following its definition \cite{Pat1969}:
\begin{equation}\label{eq:SEm}
    S^*(E)= \sigma (E) E \exp(87.21E^{-1/2} + 0.46 E)
\end{equation}

\section{\label{sec:exp} Experimental Setup}

\begin{figure}
\begin{center}
\includegraphics[width = 0.5\textwidth]{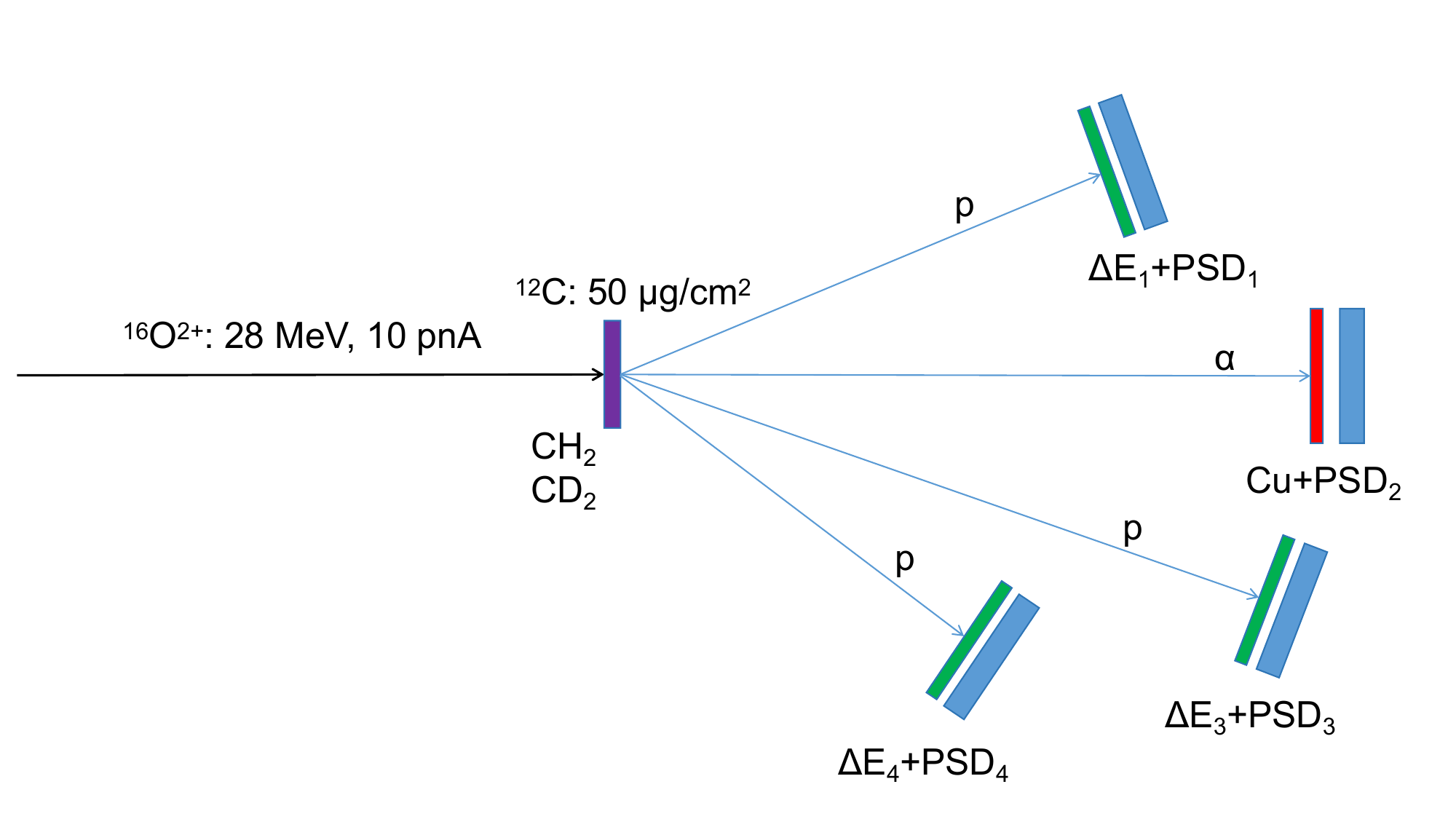}\\
\caption{(Color online) Schematic diagram of experimental setup }
\label{figSetup} 
\end{center}
\end{figure}

\begin{table}
\begin{center}
    \caption{\label{tab:detector} Layout of detectors in the experiment.}
    \begin{ruledtabular}
        \begin{tabular*}{80mm}{c@{\extracolsep{\fill}}cccc}
    Detector	& Foil	& $\rm \Delta E$	& $\rm E_r$	& Angles \\
            \hline
    det1	& 0	& 20 $\rm \mu m$ SSSD	& 500  $\rm \mu m$ PSD	& 30°±12° \\
    det2	& 12  $\rm \mu m$ Cu	& 0	& 1000  $\rm \mu m$ PSD	& 0°±8° \\
    det3	& 0	& 20 $\rm \mu m$ SSSD	& 1000  $\rm \mu m$ PSD	& -30°±8° \\
    det4	& 0	& 20  $\rm \mu m$ SSSD	& 500  $\rm \mu m$ PSD	& -60°±12° \\
      \end{tabular*}
    \end{ruledtabular}  
\end{center}
\end{table}

The experiment was performed at the HI-13 Tandem Accelerator in China Institute of Atomic Energy (CIAE). The experimental setup is shown in Fig. \ref{figSetup}. 
A carbon target, with a thickness of 50 $\rm \mu g/cm^{2}$ and a width of 1.5 mm, was bombarded by a 28 MeV $\rm ^{16}O^{2+}$ beam at an intensity of approximately 10 pnA.
The 3-body reaction $\rm ^{12}C(^{16}O, \alpha \mathit{p})^{23}Na$ was thereby induced.

The layout of detectors in the experiment is shown in Table \ref{tab:detector} and Fig. {\ref{figSetup}}.

A 1000 $\rm \mu m$ thick position-sensitive detector (PSD), shielded by a 12 $\rm \mu m$ thick copper beam-stopper foil, was positioned at \(0^{\circ} \pm 8^{\circ}\) to detect the spectator $\alpha$ particles from the quasi-free reaction. The energy and angle information was derived from the coincidence between this $0^{\circ}$ PSD and the telescope detectors.

This study introduces, for the first time in a THM experiment, the use of a beam-stopper foil to detect spectators around $0^{\circ}$, where their yield is maximized under the quasi-free mechanism. This innovation offers several critical advantages: it protects the downstream detector from damage caused by scattered heavy ions; it is transparent to light particles ($p$ and $\alpha$), enabling spectator identification; and it mitigates background from beam scattering, thereby permitting higher beam intensities. The increased beam intensity, combined with the concentrated spectator yield near $0^{\circ}$, significantly enhances the detection efficiency of quasi-free events. This leads to superior counting statistics and reduced statistical uncertainties within a given beam time.

\section{\label{sec:data} Data Analysis and Result Discussion}

\begin{figure}
\begin{center}
\includegraphics[width = 0.5\textwidth]{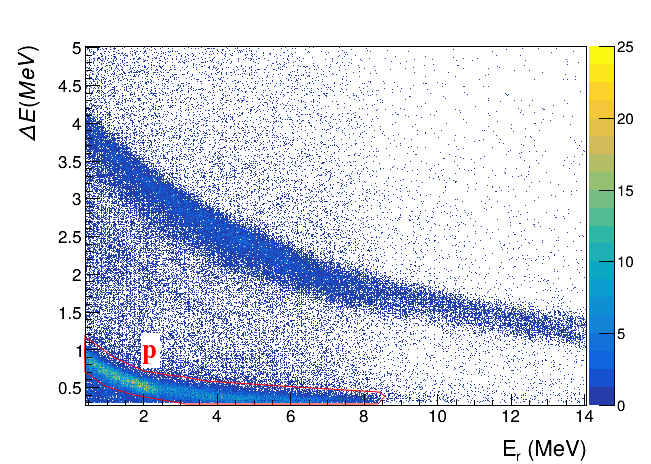}\\
\caption{(Color online) Particle identification using $\rm \Delta E-E_r$ spectrum to select $p$. }
\label{figdEE} 
\end{center}
\end{figure}

\begin{figure}
\begin{center}
\includegraphics[width = 0.5\textwidth]{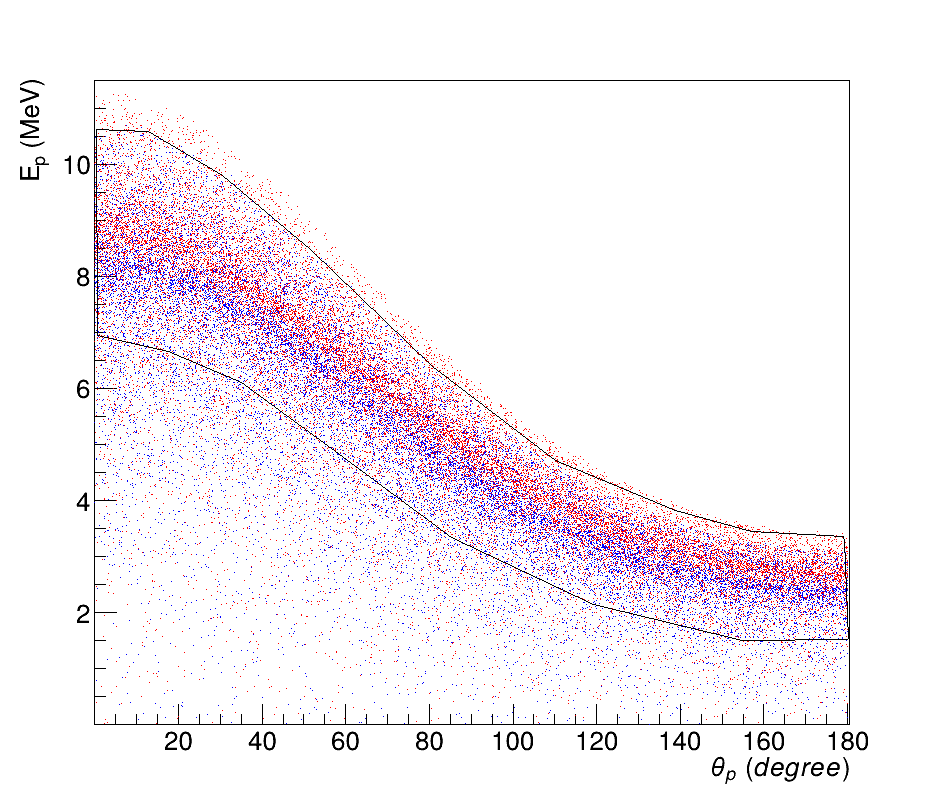}\\
\caption{(Color online) A cut (black line) is selected of the angular-energy relationship on $\rm E_p-\theta_p$ spectrum according to the simulation of the quasi-free three-body reaction events, to reduce the background from other reaction channels in experimental spectrum analysis.(Red: $p_0$ channel, Blue: $p_1$ channel.)}
\label{figcutE1th1} 
\end{center}
\end{figure}

\begin{figure}
\begin{center}
\includegraphics[width = 0.5\textwidth]{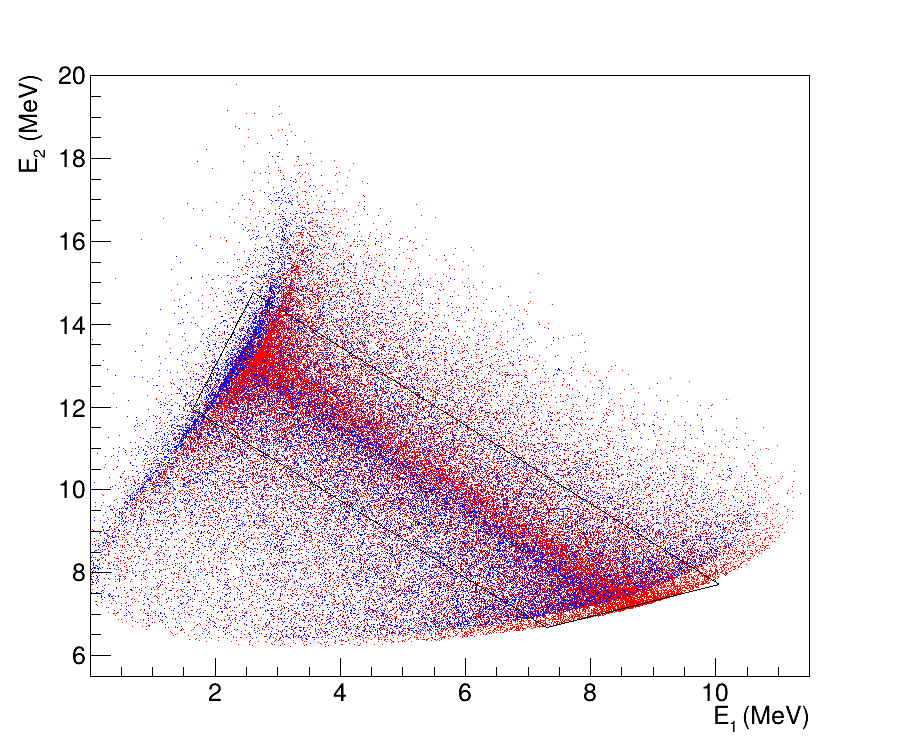}\\
\caption{(Color online)  A cut (black line) is selected of the energy relationship $\rm {E}_p-\rm{E_{^{23}Na}}$ on the $\rm {E}_1-\rm{E_2}$ spectrum according to the simulation of the quasi-free three-body reaction events, to reduce the background from other reaction channels in experimental spectrum analysis.(Red: $p_0$ channel, Blue: $p_1$ channel.)}
\label{figcutE1E2} 
\end{center}
\end{figure}

Protons were identified using the $\rm \Delta E-E_r$ spectrum (Fig. \ref{figdEE}) in coincidence with spectator $\alpha$-particles detected by the $0^\circ$ PSD. Assuming the undetected third particle to be $^{23}\mathrm{Na}$, a full kinematic reconstruction was performed to determine its energy, angle, and momentum. To suppress contributions from competing reaction channels, preliminary selection of quasi-free three-body events was implemented based on simulated correlations from a Trojan Horse method simulation. Cuts were applied to the correlation between the proton energy and angle ($\rm E_p-\theta_p$, Fig. \ref{figcutE1th1}) and the energy correlation between the proton and the reconstructed $^{23}\mathrm{Na}$ ($\rm {E}_p-\rm{E_{^{23}Na}}$, Fig. \ref{figcutE1E2}) to select quasi-free three-body events in the following steps. 

\begin{figure}
\begin{center}
\includegraphics[width = 0.5\textwidth]{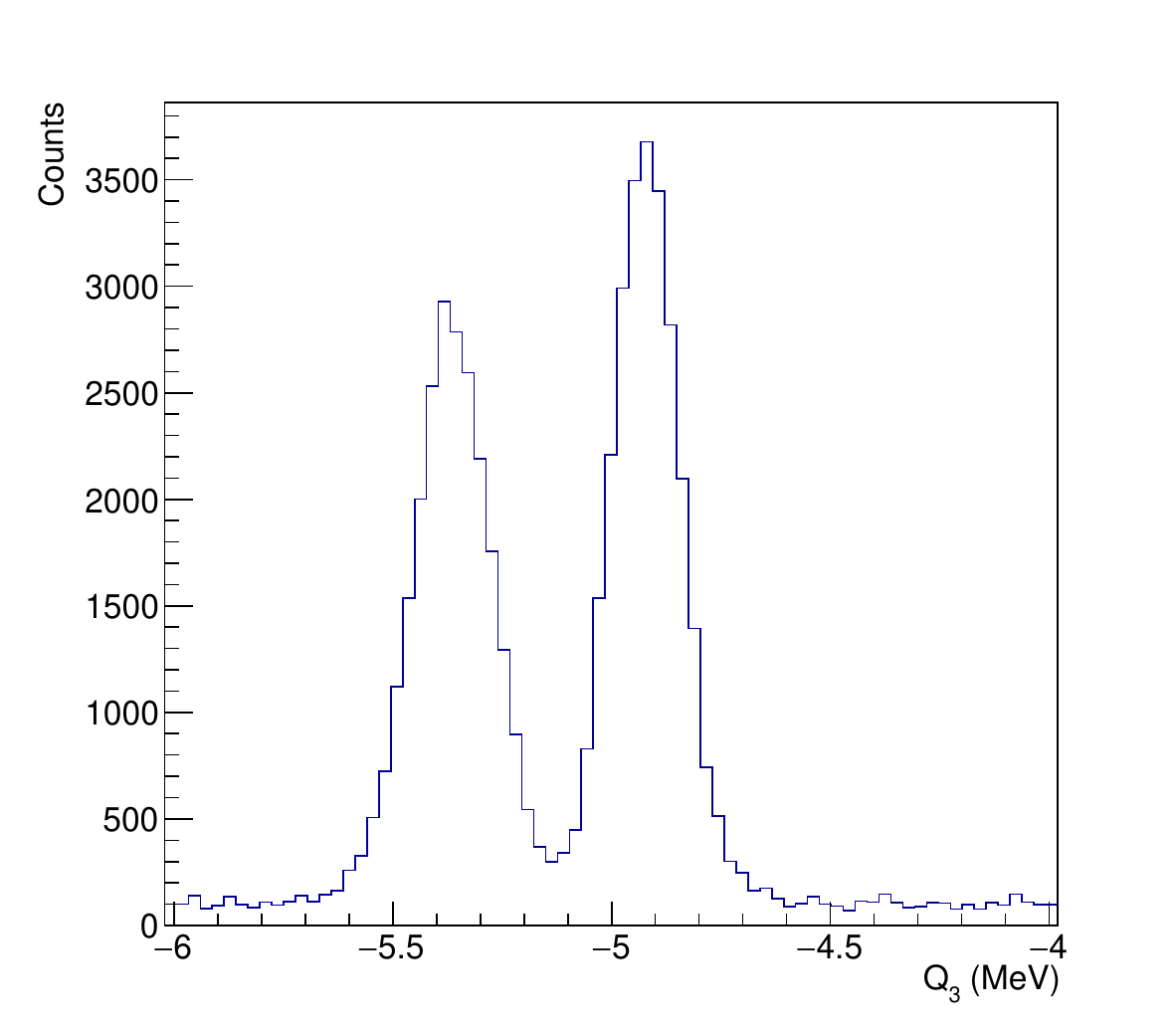}\\
\caption{(Color online) Spectrum of the experimental $Q_3$ value in the range from -6 to -4 MeV: Two distinct, partially overlapping peaks are observed on the accidental coincidence background,  corresponding to the $p_0$ channel ($Q_3$ = -4.921 MeV) and $p_1$ channel ($Q_3$ = -5.361 MeV) of the 3-body reaction $\rm ^{12}C(^{16}O, \alpha \mathit{p})^{23}Na$, respectively.}
\label{figQ3} 
\end{center}
\end{figure}

Based on kinematic reconstruction, the experimental $Q_3$ value of the 3-body reaction $\rm ^{12}C(^{16}O, \alpha \mathit{p})^{23}Na$ and the relative energy $E_{Cc}$ can be obtained.

The experimental $Q_3$ value in the range from -6 to -4 MeV is shown in Fig. {\ref{figQ3}}, under the cuts of $\rm E_p-\theta_p$ and $\rm {E}_p-\rm{E_{^{23}Na}}$ according to the simulation. 
Two distinct, partially overlapping peaks are observed on the accidental coincidence background, corresponding to the $p_0$ channel ($Q_3$ = -4.921 MeV) and $p_1$ channel ($Q_3$ = -5.361 MeV) of the 3-body reaction $\rm ^{12}C(^{16}O, \alpha \mathit{p})^{23}Na$, respectively.
Thus, the 3-body reaction events corresponding to the $p_0$ and $p_1$ channels were selected by requiring $|Q_3 + 4.92|<0.2$ MeV and $|Q_3 + 5.36|<0.2$ MeV, respectively.

To verify the dominance of the quasi-free mechanism, the momentum distribution of the spectator $\alpha$ was examined.
The momentum distribution of the spectator $\alpha$ is very sensitive to the reaction mechanism. 
This distribution is expected to reflect the internal momentum distribution of the $\alpha$ cluster within $^{16}$O only under quasi-free conditions.

\begin{figure}
\begin{center}
\includegraphics[width = 0.5\textwidth]{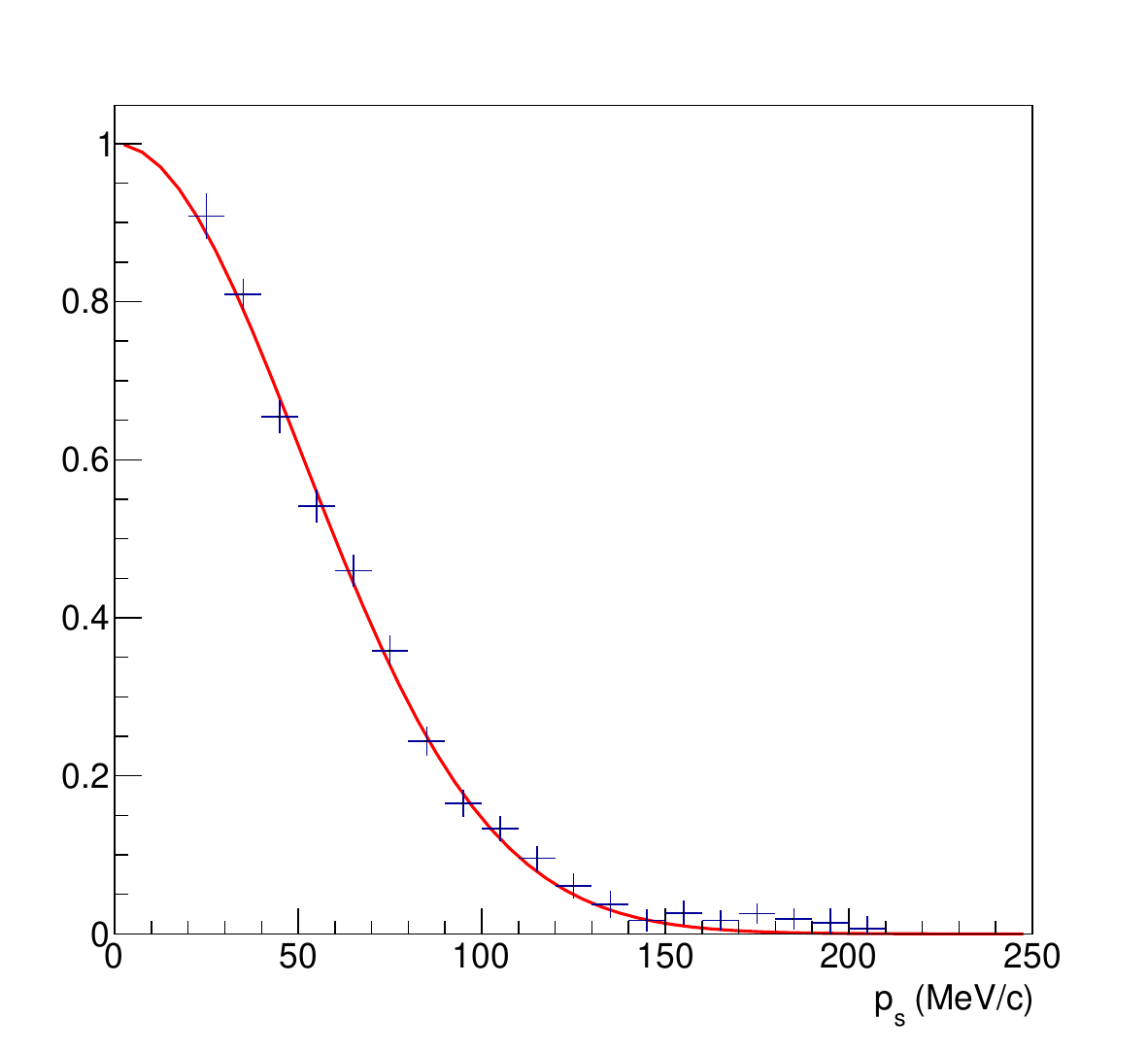}\\
\caption{(Color online) Comparison of experimental momentum distribution of spectator $\alpha$ (blue points) with the theoretical one of $\alpha$ inside $\rm ^{16}O$ (red solid line ).}
\label{figps} 
\end{center}
\end{figure}

Figure {\ref{figps}} compares, for the $p_0$ channel, the experimental momentum distribution of the spectator $\alpha$ particle in the center-of-mass system with the theoretical momentum distribution of an $\alpha$ particle inside $\rm ^{16}$O. The experimental distribution was obtained by applying three-body reaction event selection and kinematic correlations derived from THM simulations to isolate quasi-free events and eliminate interfering events as much as possible.

The solid red line represents the theoretical curve. It is essentially the Fourier transform of the coordinate-space wave function into momentum space. The wave function can be obtained by solving the Schrödinger equation with a Woods-Saxon potential (with parameters: $R$=4.65 fm, $a$ = 0.65 fm and $V_0$ = 32.45 MeV, which is obtained by adjusting it to the binding energy of 7.16 MeV for $\rm ^{16}O=^{12}C\oplus\alpha$).
This distribution exhibits a peak strictly at $p_s$=0 and follows a Gaussian-like profile for a Trojan horse nucleus with $l=0$.

The experimental data points are normalized according to the theoretical curve. 
The agreement between them in the range of $30<|p_s|<150$ (MeV/$c$) indicates the dominance of the quasi-free mechanism. 

It should be noted that the theoretical curve shown in Fig.{\ref{figps}} corresponds to the intrinsic PWIA-like momentum distribution obtained from the Fourier transform of the $\rm ^{16}O$ bound-state wave function, rather than the modified spectator momentum distribution, $|W|^2$, introduced within the DWBA framework in Eq.(\ref{eq:sec3all}). This comparison is intended as a practical consistency check. The observed good agreement indicates that distortion effects do not significantly modify the spectator momentum distribution under the selected quasi-free kinematic conditions.

\begin{figure}
\begin{center}
\includegraphics[width = 0.5\textwidth]{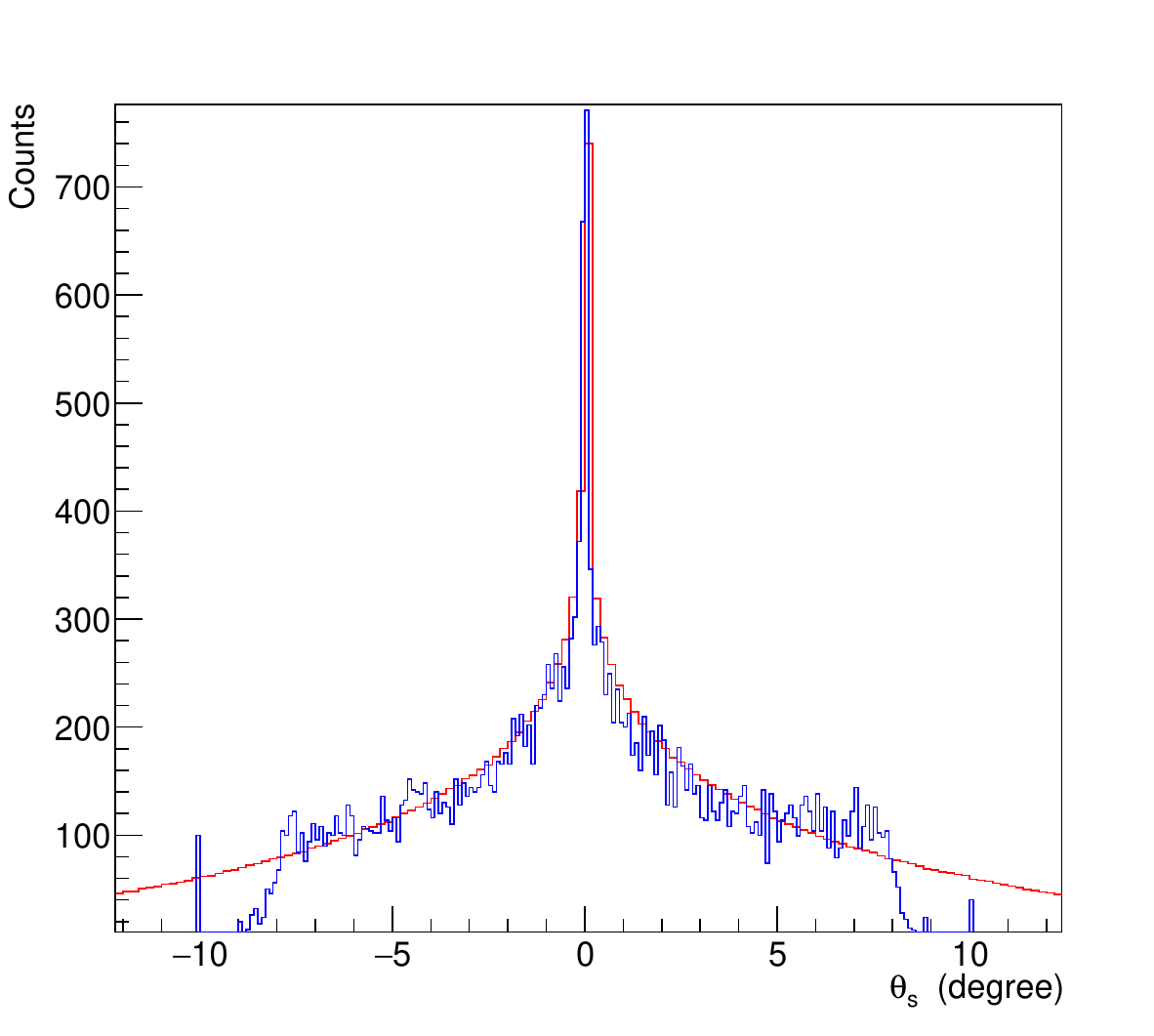}\\
\caption{(Color online) Distribution of spectator $\alpha$ around $0^{\circ}$: comparison between experimental spectrum (blue line ) and simulated result of quasi-free process (red line).}
\label{figthps} 
\end{center}
\end{figure}

Furthermore, when the Trojan horse nucleus $\rm {}^{16}O=({}^{12}C \oplus \alpha)$ is used as the beam, the spectator $\alpha$  particles are concentrated  within a small angular range around $0^{\circ}$ during the quasi-free reaction. 
As shown in Fig. {\ref{figthps}}, the angular distribution of the spectator $\alpha$ measured around $0^{\circ}$ (blue line) is obtained by selecting quasi-free three-body events as described above, and is compared with the simulation result of the THM quasi-free process (red line). The agreement indicates the contribution of the quasi-free process to the selected events.

For the subsequent data analysis, quasi-free events are selected using the cuts $30<|p_s|<150$ (MeV/$c$) and $|\theta_s|<7.0$°.

Finally, the cross section of the 2-body reaction $\rm {}^{12}C({}^{12}C,\mathit{p}){}^{23}Na$ can be extracted from the experimental data of 3-body reaction $\rm ^{12}C(^{16}O, \alpha \mathit{p})^{23}Na$ under the selection of quasi-free 3-body events as described above, according to the relationship Eq.(\ref{eq:sec3all}). 
Then, the $S^*(E)$ factor can be obtained using the definition Eq.(\ref{eq:SEm}).
The calculation shows that the high-order components of $T_l$ have little effect on the overall trend of the S factor and can be ignored. Therefore, only the main partial wave $l=0$ is considered in this work.

Due to the limited energy resolution of the experiment, combined with the limitations of the surface approximation within the DWBA framework, the resonance strength is severely suppressed and the resonance peaks are significantly broadened. This effectively washes out a large portion of the information regarding narrow resonance structures. Consequently, we were unable to extract valid information on the relevant resonances using the multi-channel R-matrix fitting code (AZURE2).
Therefore, the $S^*(E)$ obtained in this experiment, which reflects the broadening effect on the narrow resonances, can be regarded as a lower-limit reference for the actual values, and is primarily used to investigate the overall trend of $S^*(E)$ factor.

The absolute scale of the THM $S^*(E)$ factor was established by normalizing to direct experimental data in the 2.5-3.5 MeV range. The $p_0$ channel was normalized to the only direct data \cite{Maz1973}, while the $p_1$ channel was normalized to direct data sets from \cite{Spi2007,Maz1973,Agu2006,Bar2006,Fru2020,Tan2020,tan2024,nip2025}.

\begin{figure*}
\begin{center}
\includegraphics[width = 0.70\textwidth]{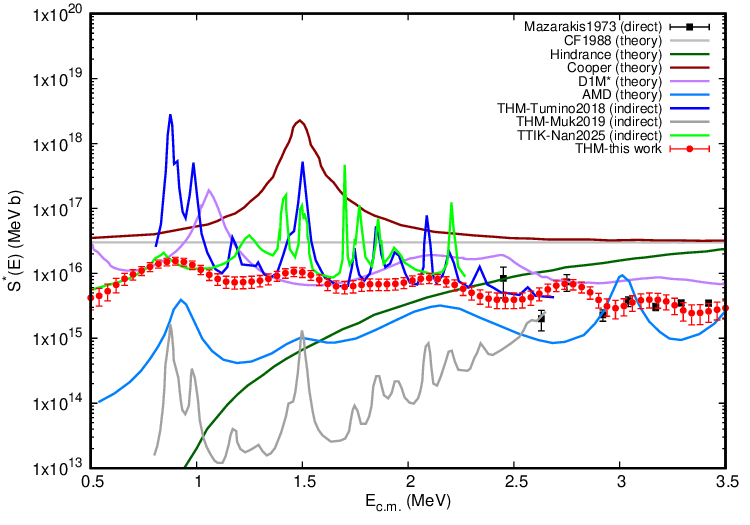}\\
\caption{(Color online) Comparison of $S^*(E)$ factor of $\rm {}^{12}C({}^{12}C,\mathit{p}_0){}^{23}Na$ extracted by THM of this work with other experimental measurements and theoretical curves. Note: the theoretical curves (CF1988, Cooper, Hindrance) represent the total $\rm S^*(E)$ factor.}
\label{figSp0} 
\end{center}
\end{figure*}

\begin{figure*}
\begin{center}
\includegraphics[width = 0.70\textwidth]{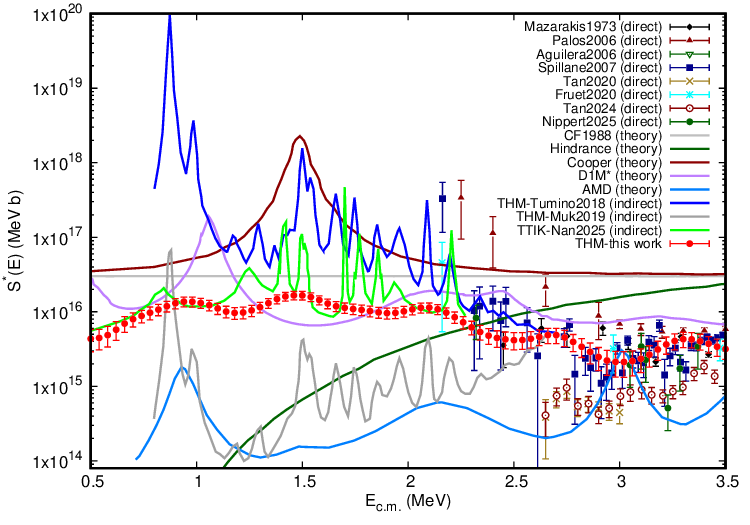}\\
\caption{(Color online) Comparison of $S^*(E)$ factor of $\rm {}^{12}C({}^{12}C,\mathit{p}_1){}^{23}Na$ extracted by THM of this work with other experimental measurements and theoretical curves. Note: the theoretical curves (CF1988, Cooper, Hindrance) represent the total $\rm S^*(E)$ factor.}
\label{figSp1} 
\end{center}
\end{figure*}

The THM-derived $S^*(E)$ factors were compared with: (i) direct measurements \cite{Maz1973, Bar2006, Agu2006, Spi2007, Tan2020, Fru2020,tan2024,nip2025}, (ii) prior THM results THM-Tumino2018 \cite{Tum2018} and the Modified-THM-Muk2019 \cite{Muk2019}, (iii) recent thick-target inverse kinematics (TTIK) data \cite{wyb2025}, and (iv) theoretical predictions \cite{CF1988, Hin2007, Cop2009, AMD2021, DIM2024}, as summarized in Figs. \ref{figSp0} and \ref{figSp1}.
The three theoretical model curves (CF1988, Cooper, Hindrance) are calculated for the total $\rm S^*(E)$ factor. Consequently, they are significantly higher than the experimental data for individual branching channels and can only serve as a qualitative guide for the overall trend in this context.

The extracted $S^*(E)$ factors for the $p_0$ and $p_1$ channels, shown in Figs. \ref{figSp0} and \ref{figSp1}, exhibit broad resonant structures in the 0.5 to 3.0 MeV range, with peaks located near 0.8, 1.5, 2.1, and 2.7 MeV. In particular, the resonance near the Gamow energy ($\approx 1.5$ MeV) is confirmed, as suggested by the Cooper prediction \cite{Cop2009} and found in the indirect experiment by Tumino2018 \cite{Tum2018}.

However, due to the limited energy resolution of the experimental detection system and the approximations in the DWBA theoretical model, the resonance widths are significantly broader than those with R-matrix fitting in Tumino2018 \cite{Tum2018} and Nan2025 \cite{wyb2025}.
The original narrow resonance is broadened, or many narrow resonances are superimposed and merge into a single broader peak.

Without considering the resonance details of $S^*(E)$, the overall trend of our experimental results is in reasonable agreement with the THM-Tumino2018\cite{Tum2018} and TTIK2025 data\cite{wyb2025}, but differs significantly from the trend of the Coulomb-corrected data THM-Muk2019\cite{Muk2019}.
The THM-Muk2019 data generally exhibit a trend where resonance structures are superimposed on a non-resonant background that decreases with decreasing energy. In particular, within the energy range of 1.2–2.5 MeV, the variation trends show significant differences.

Although our results appear to be relatively closer to the TTIK data in the very low energy region (0.5–1.1 MeV), the resonance peaks are significantly broadened due to the limited energy resolution of this experiment and the limitation of the surface approximation within the DWBA framework. Therefore, our results can only serve as a lower-limit reference for the actual values, reflecting the overall trend, and should not be over-interpreted as a definitive judgment between different datasets.

In comparison with theoretical models, our data exhibit a resonant peak near 1.5 MeV, in agreement with the prediction of Cooper \cite{Cop2009}, and distinct from those of CF1988 \cite{CF1988} and the Hindrance model \cite{Hin2007}.
Although our results appear to be relatively close to the D1M* curve \cite{DIM2024}, the agreement in the positions of the resonant peaks is poor. In contrast, our results show a modest agreement with the AMD model \cite{AMD2021} in terms of the resonance structures in the energy range of 0.5-2.5 MeV. Nevertheless, the AMD model underestimates the overall magnitude, as its calculations do not account for non-resonant contributions or possible interference effects.

We find no evidence of a hindrance effect \cite{Hin2007} in the energy range above 1 MeV.
Although the  $S^*(E)$ factor exhibits a gradual decline below 0.8 MeV, this behavior is primarily attributed to the downward tail of the resonance peak located near 0.8 MeV.
Moreover, the hindrance model predicts that the onset of the hindrance effect occurs at approximately 3.5 MeV, leading to a sharp decrease in the  $S^*(E)$ factor as the energy is reduced further.
Consequently, the observed low-energy trend cannot be interpreted as definitive evidence for a hindrance effect.

Additionally, the $S^*(E)$ results from our recent work on the alpha emission channels of the $\rm ^{12}C+^{12}C$ reaction \cite{lcbplb} show similar behavior to the proton emission channels.

The energy resolution of the experimental setup is the key factor limiting the accuracy of the results in this work. 
The overall energy resolu­tion is estimated to be about $\sigma$= 96 keV in the center-of-mass system, corresponding to FWHM = 226 keV.
A comprehensive evaluation suggests that non-uniformities in the beam-stopper foil thickness were likely the primary contributor to the degraded energy resolution.

\section{\label{sec:sum} Summary}

In summary, resonances in the $\rm ^{12}C+^{12}C$ reaction within the Gamow window play a decisive role in stellar carbon burning and explosive nucleosynthesis. However, they remain inaccessible to precise direct measurements due to extreme Coulomb suppression at astrophysical energies.

In this work, we performed an indirect measurement of the $\rm {}^{12}C(^{16}O,\alpha \mathit{p}){}^{23}Na$ reaction at CIAE, employing $\rm {}^{16}O=({}^{12}C \oplus \alpha)$ as the Trojan horse nucleus due to its lower binding energy, which favors the quasi-free reaction mechanism.

A key experimental innovation was the introduction of a copper beam-stopper foil — the first such implementation in a THM study — enabling the detection of spectator particles near $0^\circ$, where their yield is maximized under quasi-free kinematics.

The astrophysical $S^*(E)$ factor for the $\rm {}^{12}C({}^{12}C,\mathit{p}){}^{23}Na$ reaction was extracted using the DWBA-based THM formalism, without R-matrix fitting.

Our results confirm the presence of resonant structures within the Gamow window, notably around 1.5 MeV.

We find no evidence for a hindrance effect \cite{Hin2007} in the measured energy range. 

Without considering the resonance details of $S^*(E)$, the overall trend of our results is qualitatively in reasonable agreement with the THM-Tumino2018\cite{Tum2018} and TTIK2025 data\cite{wyb2025}, but differs significantly from the Coulomb-corrected data THM-Muk2019\cite{Muk2019}. 

Although our results appear to be relatively closer to the TTIK data within 0.5–1.1 MeV, the resonances are significantly broadened due to the limited experimental energy resolution and the surface approximation in the DWBA framework. Therefore, our results can only serve as a lower-limit reference for the actual $S^*(E)$ values, reflecting the overall trend, and should not be over-interpreted as a definitive judgment between different datasets.

This work highlights the power of the THM combined with near- $0^\circ$ spectator detection for probing key astrophysical reactions. It provides crucial experimental constraints on the low-energy behavior of carbon–carbon fusion, essential for refining models of stellar evolution and explosive nucleosynthesis.

\begin{acknowledgments}
We thank the Nuclear Reaction Group in CIAE for their kind help during the experimental preparation and measurement.  
We would like to thank the staff of the HI-13 tandem accelerator laboratory for providing the experimental beam and targets.
We also thank Prof. Weiping Liu, Prof. Xiaodong Tang, Prof. Zhihong Li, Prof. Bing Guo, Prof. Youbao Wang, and Prof. Shuhua Zhou for helpful discussions on the research work. 
\end{acknowledgments}

\bibliography{THM-OCp-PRC}

\end{document}